\documentclass[a4paper,11pt]{article}
\usepackage{jinstpub}
\usepackage{lineno}
\usepackage{smartdiagram}
\usepackage{graphicx}
\usepackage{anyfontsize,color}
\usepackage{tikz}
\usepackage{verbatim}
\usepackage{amsmath}
\usepackage{amsfonts}

\usepackage{amssymb}
\usepackage{marvosym}
\usepackage{float}
\usepackage[section]{placeins}
\usetikzlibrary{arrows,shapes}
\usetikzlibrary{positioning, arrows.meta, shapes.geometric,fit,backgrounds,shapes.misc}
\usetikzlibrary{decorations.pathmorphing}
\usepackage{lmodern}
\usepackage{makeidx}
\usepackage{wrapfig}
\usepackage{adjustbox}
\usepackage{tcolorbox}
\usepackage{metalogo}
\usepackage{multimedia,animate}
\usepackage{media9}
\usepackage{beton}
\usepackage{euler}
\usepackage{enumitem}
\usepackage[T1]{fontenc}

\tikzset{
    vertex/.style = {
        circle,
        fill=black,
        outer sep=2pt,
        inner sep=1pt,
    }
}

\newcommand{\rat}[2]{\mbox{$#1$/$#2$}}

\title{{\boldmath Hybrid Machine-Learning Particle Identification for the ePIC Proximity-Focusing RICH}}

\author[a,1]{D.~H.~Dongwi,}
\author[a,2]{C.-J.~Na\"im,}
\author[a,3]{L.~Rhode,}
\author[a]{A.~Deshpande}

\affiliation[a]{Center for Frontiers in Nuclear Science (CFNS), 
Department of Physics and Astronomy, 
Stony Brook University, 
Stony Brook, NY 11794, USA}

\note[1]{dongwi.dongwi@stonybrook.edu}
\note[2]{charlesjoseph.naim@stonybrook.edu}
\note[3]{lucas.rhode@stonybrook.edu}

\abstract{
We present a machine-learning–based particle-identification study for the proximity-focusing Ring Imaging Cherenkov (pfRICH) detector of the ePIC experiment at the Electron–Ion Collider. Operating in the backward region ($-3.5 \lesssim \eta \lesssim -1.5$), the pfRICH is designed to provide at least $3\sigma$ separation among pions, kaons, and protons up to 7~GeV/$c$ for Semi-Inclusive Deep Inelastic Scattering measurements. Using a standalone \textsc{Geant4} simulation of the pfRICH, we develop a hybrid model that combines convolutional neural network–based feature extraction with gradient-boosted decision-tree classifiers. This approach establishes a controlled single-particle benchmark for Cherenkov-ring pattern recognition and particle separation, and introduces a hybrid CNN + gradient-boosting framework for next-generation Cherenkov detectors at the EIC. In this sanitized sample the kinematic and detector-level features already provide essentially all of the discrimination. The image-based component is retained as a proof of concept, to be evaluated under realistic occupancy and background conditions.
}


\begin{document}
\maketitle
\flushbottom

\section{Introduction}
\label{sec:intro}

\begin{figure}
    \centering
    \includegraphics[scale=0.42]{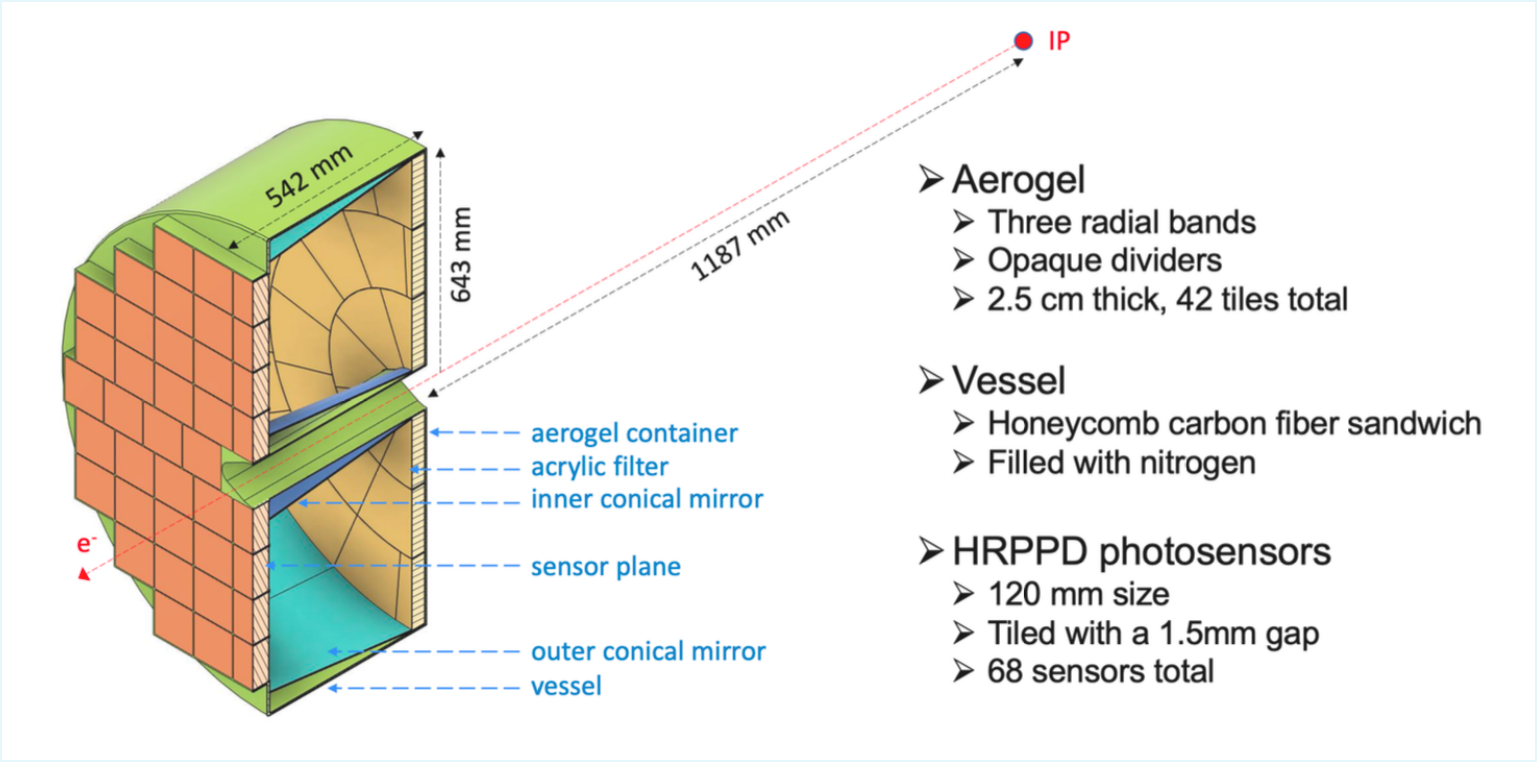}
    \caption{Schematic view of the proximity-focusing Ring Imaging Cherenkov (pfRICH) detector.}    
    \label{fig:pfrich_detector}
\end{figure}

The proximity-focusing Ring Imaging Cherenkov (pfRICH)~\cite{Chatterjee:2024zrn} serves as a Particle Identification (PID) detector in the backward region, $-3.5 \lesssim \eta \lesssim -1.5$, of the upcoming ePIC experiment at the Electron–Ion Collider (EIC)~\cite{AbdulKhalek:2021gbh}. The ability to distinguish final-state hadron production in electron–nucleus collisions is crucial for Semi-Inclusive Deep-Inelastic Scattering (SIDIS) measurements. The pfRICH consists of three main components: an aerogel radiator, a mirror system, and a photon sensor plane as illustrated in Fig.~\ref{fig:pfrich_detector}.
Charged particles traversing the aerogel emit Cherenkov photons, which are reflected by mirrors and projected onto the photon sensor plane, forming characteristic ring patterns. The measurement of the ring radius allows reconstruction of the Cherenkov angle, which is directly related to the velocity of the particle. The Cherenkov emission angle is defined as
\begin{equation}
    \cos{\left(\theta_{c}\right)} = \frac{1}{n\beta},
    \label{eq:cherenkov}
\end{equation}
where $n$ is the refractive index of the aerogel and $\beta = \frac{v}{c}$ is the particle’s velocity in units of the speed of light. For relativistic particles ($\beta \lesssim 1$), expanding to first order in \rat{m^2}{p^2} gives
\begin{equation}
    \theta_{c}^{\,2} \simeq \theta_{\mathrm{sat}}^{\,2}
    - \frac{1}{n}\frac{m^{2}}{p^{2}},
    \label{eq:thetaapprox}
\end{equation}
where $\theta_{\mathrm{sat}}$ is the saturation angle corresponding to ultra-relativistic emission.

\begin{figure}
    \centering
    \begin{minipage}{.5\textwidth}
        \centering
    \includegraphics[scale=0.32]{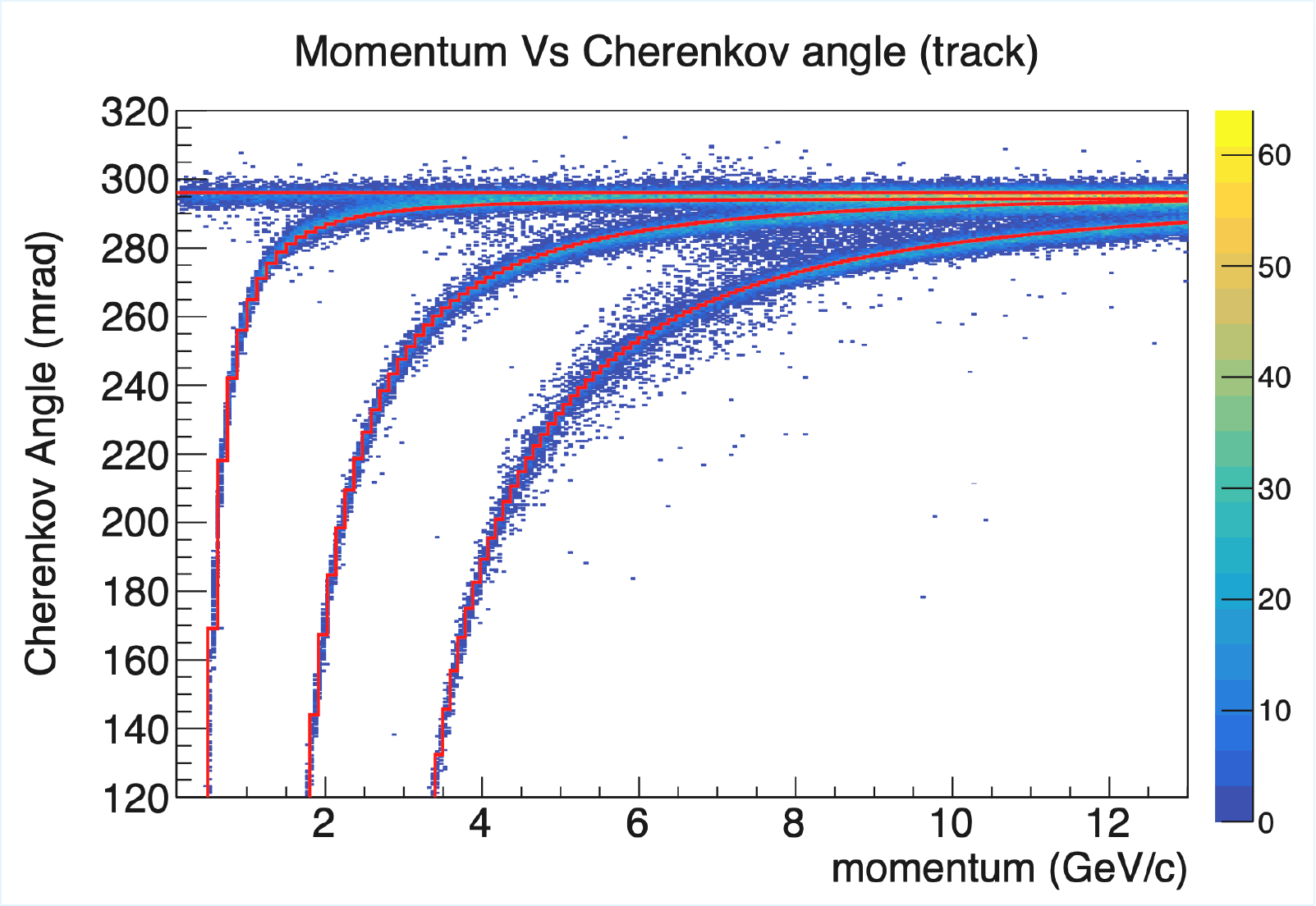}
    \end{minipage}%
    \begin{minipage}{0.5\textwidth}
        \centering
    \includegraphics[scale=0.32]{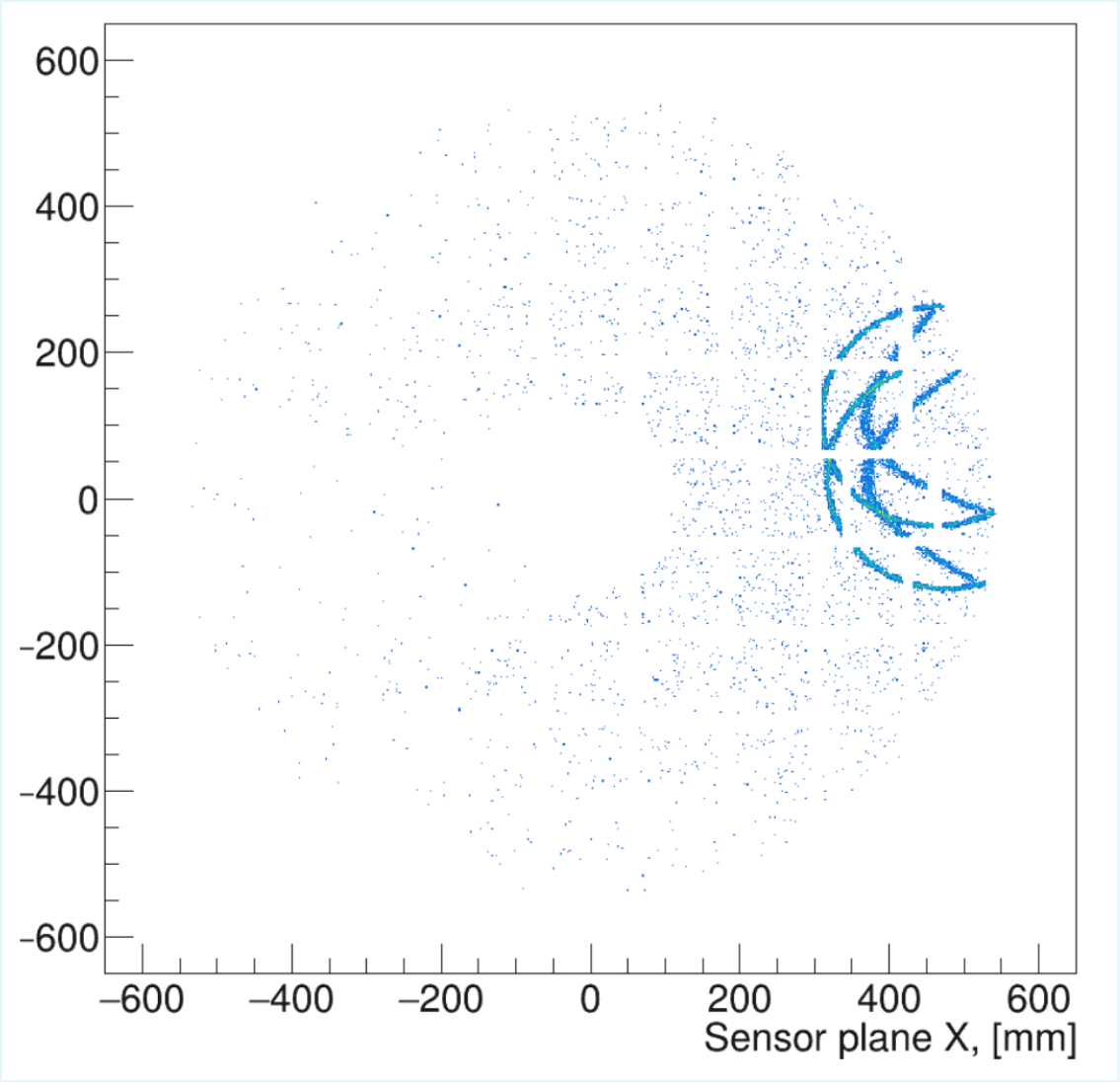}
    \end{minipage}
     \caption{Left: Reconstructed Cherenkov angle (mrad) as a function of hadron's momentum (GeV/c) for electrons, pions, kaons and protons using Eq.~\eqref{eq:cherenkov}. Right: Overlapping reconstructed rings of pion and kaon particles on the sensor plane.}
    \label{fig:cherenkov_rings}
\end{figure}

This implies that for a given momentum measured by the tracking system, the Cherenkov angle $\theta_{c}$ depends on the particle mass (see Fig.~\ref{fig:cherenkov_rings}). This property allows the pfRICH detector to separate pions, kaons, and protons by comparing the measured angle with the expected values. The pfRICH is expected to provide at least $3\sigma$ separation up to 7~GeV/$c$ \cite{Chatterjee:2024zrn}.

Conventional RICH particle identification typically proceeds by reconstructing the Cherenkov angle or ring radius associated with a charged-particle track and comparing the measured response with the expectations for different mass hypotheses. In likelihood-based approaches, the probability of the observed photon-hit pattern is evaluated under each particle hypothesis using the expected Cherenkov angle, photon yield, detector resolution, and background occupancy. Such methods provide a natural physics baseline for RICH PID. However, their performance can become more challenging in the presence of incomplete rings, nonuniform detector response, overlapping rings from multiple tracks, and optical or electronic backgrounds. These effects motivate complementary pattern-recognition approaches that can exploit the full spatial hit distribution on the sensor plane.

Beyond standard reconstruction methods, machine-learning techniques are explored in order to improve the pfRICH particle-identification performance. The Cherenkov ring patterns recorded on the photon sensor plane form a natural image-like structure that can be exploited by using advanced pattern-recognition algorithms. In the next section, we introduce a hybrid model that combines pattern recognition with boosted decision trees, using both hadron kinematic information and hit-level features to improve pfRICH PID performance.

\section{Simulation and data}
\label{sec:simulation}

The datasets used in this study are produced using a standalone \textsc{Geant4}-based pfRICH simulation~\cite{eic_pfrich_v110}, which provides a detailed description of the aerogel radiator, mirror system, and photosensor response. Single-particle samples are generated for electrons ($e^{-}$), pions ($\pi^{-}$), kaons ($K^{-}$) and protons ($p$) over a momentum range of $1$--$12$~GeV/$c$ and pseudorapidities $-3.5 \leq \eta \leq -1.5$, with uniformly distributed azimuthal angles. The simulation is performed on a structured $(p,\eta)$ grid to ensure uniform coverage of the pfRICH acceptance.

This study uses a simplified single-particle setup rather than a full ePIC event simulation. Each simulated event contains one primary charged particle traversing the pfRICH acceptance, and the training labels correspond to the generated primary particle species. Consequently, the results reported here should be interpreted as a controlled detector-response and pattern-recognition study. Effects from realistic event multiplicity, overlapping rings from multiple tracks, beam-related backgrounds, and full detector occupancy are not included at this stage and will be addressed in future studies within the full ePIC reconstruction framework.

\begin{figure}
    \centering
    \includegraphics[width=\textwidth]{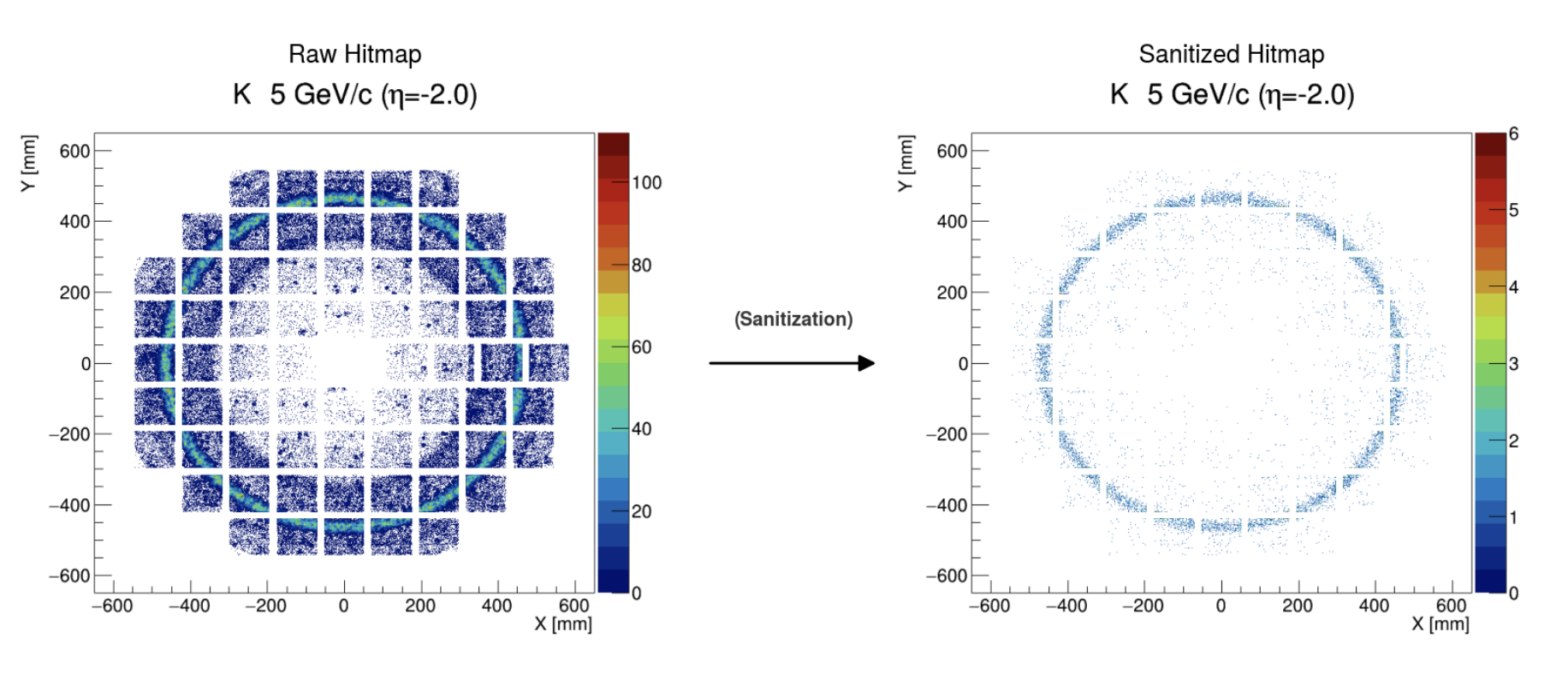}
    \caption{Comparison of photon hit patterns on the pfRICH sensor plane before and after preprocessing. Left: raw event, including background contributions. Right: corresponding sanitized event used for machine-learning training. The sanitization procedure removes low-energy hits ($<1$~eV) and applies fiducial cuts to isolate the primary Cherenkov signal.}
    \label{fig:hitmaps}
\end{figure}

To prepare the raw detector output for machine-learning training, a sanitization and quality-assurance (QA) procedure is applied to the \textsc{Geant4} hit records. Only primary particles corresponding to the target species ($e^{-}$, $\pi^{-}$, $K^{-}$, $p$) are retained in order to suppress secondary particles originating from material interactions. Hits with energies below 1~eV and those outside the active fiducial sensor area are removed to reject optical background and dark noise. The resulting sanitized hitmaps (Fig.~\ref{fig:hitmaps}) isolate the primary Cherenkov signal and provide clean, sparse patterns suitable for pattern-recognition models.

This sanitization produces a clean sample that is useful for validating the reconstruction and classification pipeline, but it also removes several effects that are expected in realistic data. The results presented below should therefore be interpreted as performance under controlled single-particle conditions rather than as a final estimate of the full experimental PID performance.

For context, the pfRICH design produces approximately
$\langle N_{\mathrm{pe}} \rangle \sim 12$ detected Cherenkov photons
per saturated track from the aerogel ring, plus an additional cluster of
$\sim 80$ photons per track from the fused-silica window of each HRPPD
photosensor used for the time-of-flight
measurement~\cite{Chatterjee:2024zrn}. Both populations are visible in
the present simulation as a clearly bimodal $n_{\mathrm{hits}}$
distribution for saturated electrons, with peaks consistent with these
design values. The pfRICH also operates in a low-occupancy regime: in
SIDIS events with inelasticity $y > 0.7$, approximately $55\%$ of events
have no additional charged particle, $30\%$ have one, and $10\%$ have
two charged particles in proximity to the scattered
electron~\cite{Chatterjee:2024zrn}, so overlapping rings from multiple
tracks are uncommon but not negligible.

A radial acceptance cut of $R < 650$~mm is applied to preselect hits within the active sensor plane. The final sanitized datasets contain 2,260,869 events distributed across the four particle species: 625,812 electrons, 568,140 pions, 528,816 kaons, and 538,101 protons.

\begin{figure}
    \centering
    \includegraphics[width=\textwidth]{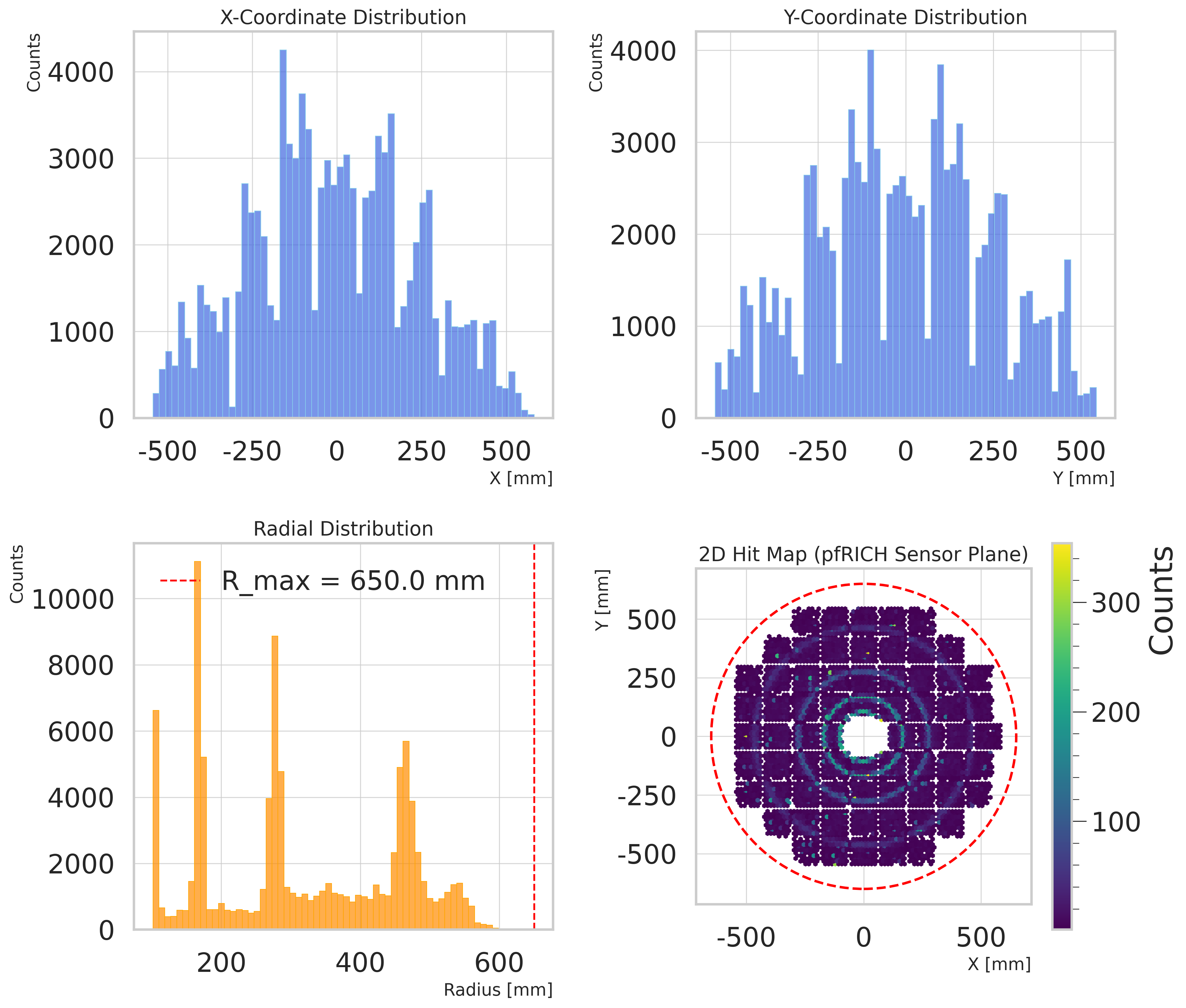}
    \caption{Quality-assurance distributions for sanitized photon hits, aggregated over many single-particle events. Top: global sensor-plane $x$ and $y$ coordinate distributions, showing the segmented photosensor structure and inactive regions between tiles. Bottom left: distribution of the global hit radius $R_{\mathrm{global}}=\sqrt{x^2+y^2}$ measured from the detector coordinate origin. This is a fiducial-acceptance and occupancy diagnostic after the full optical transport, not the local Cherenkov ring radius used for PID. The red dashed line indicates the fiducial acceptance boundary $R_{\mathrm{max}}=650$~mm. Bottom right: integrated two-dimensional hit distribution on the pfRICH sensor plane, including the central aperture corresponding to the beam pipe.}
    \label{fig:data_qa}
\end{figure}

Quality-assurance distributions of the sanitized datasets are presented in Fig.~\ref{fig:data_qa}. The $x$ and $y$ coordinate distributions (top panels) exhibit the segmented structure of the photosensor array, with gaps corresponding to inactive regions between sensor tiles. The radial distribution (bottom left) shows distinct peaks associated with Cherenkov rings at different momenta, with all hits contained within the $R_{\mathrm{max}} = 650$~mm acceptance. The two-dimensional hitmap (bottom right) displays the integrated ring pattern on the pfRICH sensor plane, including the central aperture corresponding to the beam pipe.

\begin{figure}
    \centering
    \includegraphics[width=0.90\textwidth]{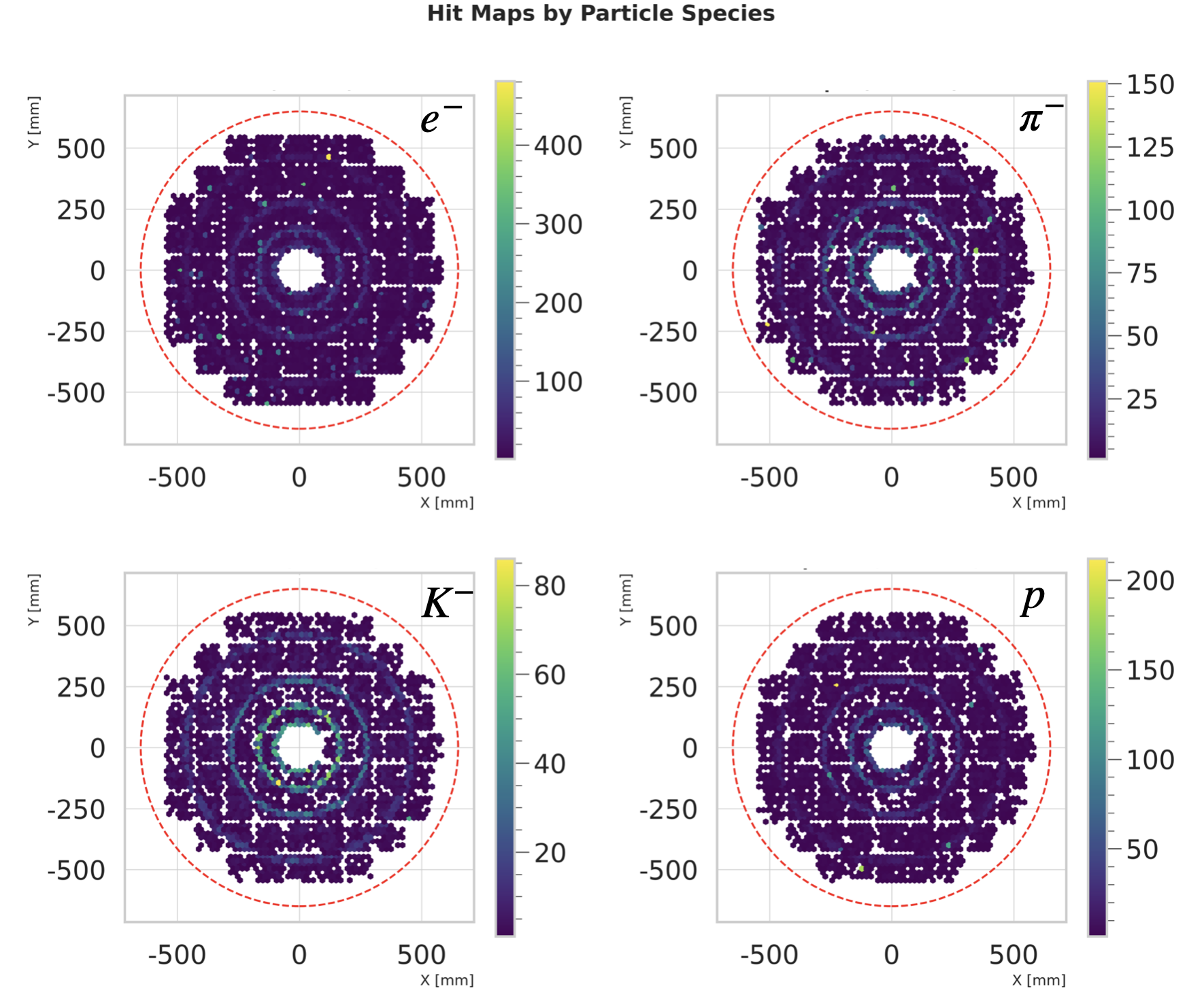}
    \caption{Aggregated hit maps on the pfRICH sensor plane for each particle species, integrated over the simulated momentum and pseudorapidity range. The plots are shown in global sensor-plane coordinates after the full pfRICH optical transport, including direct and mirror-reflected photon paths. The visible radial structures therefore reflect the detector geometry, optical-path configuration, and integration over kinematics, and should not be interpreted as fixed-momentum local Cherenkov ring radii. The red dashed circle indicates the fiducial acceptance boundary $R_{\mathrm{max}}=650$~mm. The species-dependent radial structure shown here is specific to the discrete $(p,\eta)$ sampling grid in global sensor-plane coordinates and is illustrative of the qualitative detector response; integrated over a continuous, physical $\eta$ distribution this global-coordinate structure would largely smear out. It is not the local Cherenkov-angle observable used for PID (cf.~Fig.~\ref{fig:cherenkov_rings}).}
    \label{fig:hitmaps_by_particle}
\end{figure}

The hit coordinates shown in Figs.~\ref{fig:data_qa} and~\ref{fig:hitmaps_by_particle} are global sensor-plane coordinates after the full pfRICH optical transport. The radial variable shown below is therefore
$R_{\mathrm{global}}=\sqrt{x^2+y^2}$, measured relative to the detector coordinate origin. This quantity is used as a detector-occupancy and fiducial-acceptance diagnostic and should not be interpreted as the local Cherenkov ring radius. In the pfRICH reconstruction, the relevant observable is the Cherenkov angle, obtained from the photon detection point, the emission point, and the charged-particle direction. The optical transport includes both direct photon paths and paths reflected by the conical mirrors, so the apparent radius of structures in global sensor-plane coordinates cannot be inferred from a simple proximity-gap estimate alone.

The aggregated hit patterns separated by particle species are shown in Fig.~\ref{fig:hitmaps_by_particle}. Electrons ($e^{-}$) produce the largest rings, corresponding to the saturation angle $\theta_{\mathrm{sat}}$ since $\beta \approx 1$ across the full momentum range. Because the maps aggregate single-particle events with uniformly distributed azimuth, the individual patterns superimpose into axially symmetric bands centered on the detector axis. The apparent common center reflects this azimuthal averaging rather than a common track impact point. Pions, kaons, and protons produce progressively smaller rings at the same momentum, reflecting their larger masses and correspondingly smaller Cherenkov angles according to Eq.~\eqref{eq:thetaapprox}. The clear ring structure observed in these aggregated hitmaps motivates the use of geometric features, such as the ring radius, together with pattern-recognition techniques for particle identification.

The datasets are divided into training and test sets using a stratified random split at the event level. This stratification guarantees that each particle species ($e^{-}$, $\pi^{-}$, $K^{-}$, $p$) is proportionally represented in both subsets, thereby preventing class imbalance from biasing the evaluation. A fixed random seed is used to ensure reproducibility.

\section{Machine learning model}
\label{sec:model}

The hybrid model combines a convolutional neural network (CNN) encoder with an \texttt{XGBoost} gradient-boosted decision-tree classifier~\cite{Chen_2016}. The pipeline consists of two main stages:
\begin{enumerate}
    \item a CNN encoder for extracting spatial features from per-event photon hit patterns, and
    \item an \texttt{XGBoost} classifier for particle identification.
\end{enumerate}
Together, these components form a multi-stage pipeline optimized for combining image-like detector information with tabular kinematic inputs, as illustrated in Fig.~\ref{fig:hybridmodel_flowchart}.

\begin{figure}[H]
\centering
\begin{tikzpicture}[
    >=latex,
    process/.style={rectangle, draw, thick, rounded corners,
                    minimum width=5.0cm, minimum height=0.8cm, align=center},
    arrow/.style={->, thick, shorten >=1pt, shorten <=1pt},
  ]
  \node[process] (step1) {\textbf{I.} Per-event histogram construction \\
  Build $64\times64$ TH2 images from photon hits};
  \node[process, below=0.4cm of step1] (step2)
  {\textbf{II.} CNN feature extraction \\
  Encode images $\rightarrow$ latent embeddings $\mathbf{z}$};
  \node[process, below=0.4cm of step2] (step3)
  {\textbf{III.} Feature combination \\
  Concatenate $\mathbf{z}$ with beam particle features};
  \node[process, below=0.4cm of step3] (step4)
  {\textbf{IV.} XGBoost training \\
  $f_{\mathrm{xgb}}(\mathbf{z}, \mathbf{x}_{\mathrm{beam}})$};
  \node[process, below=0.4cm of step4] (step5)
  {\textbf{V.} Inference pipeline \\
  Histogram $\rightarrow$ CNN $\rightarrow$ XGBoost $\rightarrow$ PID};
  \draw[arrow] (step1) -- (step2);
  \draw[arrow] (step2) -- (step3);
  \draw[arrow] (step3) -- (step4);
  \draw[arrow] (step4) -- (step5);
\end{tikzpicture}
\caption{Algorithm flow of the hybrid CNN+XGBoost model, illustrating per-event histogram construction, CNN feature extraction, feature combination, and classifier inference.}
\label{fig:hybridmodel_flowchart}
\end{figure}

\subsection{Pattern recognition}

The CNN encoder processes per-event photon hit patterns to extract spatial features that capture the geometric structure of Cherenkov rings. For each event, photon hits are binned into a $64\times64$ histogram representing the spatial distribution on the sensor plane. The CNN architecture consists of:
\begin{enumerate}
    \item \textbf{Convolutional layers:} three Conv2D layers with 32, 64, and 128 filters, using ReLU activation and max-pooling to extract hierarchical spatial features;
    \item \textbf{Global average pooling:} reducing the spatial dimensionality while preserving the learned feature content;
    \item \textbf{Dense projection:} mapping the pooled features to a 64-dimensional latent embedding $\mathbf{z}$ with layer normalization.
\end{enumerate}
The CNN encoder is not trained end-to-end with the classifier. Instead, the latent embeddings are computed separately and subsequently combined with kinematic features for classification.

\subsection{XGBoost}

The \texttt{XGBoost} classifier operates on a 72-dimensional input vector, composed of eight kinematic and detector-level features augmented by the 64-dimensional CNN latent representation. The kinematic and detector features include the transverse momentum $p_T$, total momentum magnitude $P_{\mathrm{total}} = |\vec{p}|$, pseudorapidity $\eta$, polar and azimuthal angles $(\theta_{\mathrm{beam}}, \phi)$, the reconstructed Cherenkov angle $\theta_c$, the mean photon arrival time $T_{\mathrm{hits}}$, and the number of detected photon hits $n_{\mathrm{hits}}$. These quantities are concatenated with the CNN embedding $\mathbf{z}$ to form the input to the gradient-boosted classifier.

For the multiclass classifier used for the main PID results, the assigned particle species is taken as the class label returned by the XGBoost \texttt{multi:softmax} objective. This corresponds to selecting the particle hypothesis with the largest classifier score among the four target species. The efficiencies and mis-identification probabilities reported below are computed from these assigned class labels.

\section{Results}
\label{sec:results}

The performance of the hybrid CNN+XGBoost model is evaluated on an independent test set of 570,963 events, while 1,712,886 events are used for training. The classifier operates on a 72-dimensional feature vector obtained by combining 64-dimensional CNN embeddings extracted from per-event hit patterns with eight beam particle features ($P_{\mathrm{total}}$, $p_T$, $\eta$, $\theta_{\mathrm{beam}}$, $\phi$, reconstructed Cherenkov angle, mean photon arrival time, and number of detected photon hits).

\subsection{Model performance}
\label{sec:performance}

The XGBoost classifier is trained using 300 boosting rounds, a maximum tree depth of 6, and a learning rate of 0.1. Table~\ref{tab:hybrid_metrics} summarizes the overall classification performance of the hybrid model.

\begin{table}[H]
\centering
\begin{tabular}{lcccc}
\hline
\textbf{Model} & \textbf{Accuracy} & \textbf{F1-score} & \textbf{Precision} & \textbf{Recall} \\
\hline
CNN+XGBoost & 95.07\% & 0.951 & 0.951 & 0.951 \\
\hline
\end{tabular}
\caption{Overall classification metrics for the hybrid CNN+XGBoost model using combined CNN embeddings and beam particle features.}
\label{tab:hybrid_metrics}
\end{table}

Table~\ref{tab:perclass_hybrid} reports the per-class identification performance. Protons achieve the highest efficiency (98.3\%), reflecting their distinct Cherenkov signature at lower $\beta$ values. Kaons, which are particularly challenging for traditional reconstruction due to their intermediate mass, reach an efficiency of 96.6\% with the hybrid approach. Electrons and pions achieve efficiencies of 93.3\% and 94.1\%, respectively.

\begin{table}[H]
\centering
\begin{tabular}{lccc}
\hline
\textbf{Particle} & \textbf{Precision} & \textbf{Recall} & \textbf{F1-score} \\
\hline
$e^-$     & 0.972 & 0.933 & 0.952 \\
$\pi^-$   & 0.918 & 0.941 & 0.929 \\
$K^-$     & 0.938 & 0.966 & 0.952 \\
$p$       & 0.991 & 0.983 & 0.987 \\
\hline
\end{tabular}
\caption{Per-class classification metrics for the hybrid CNN+XGBoost model.}
\label{tab:perclass_hybrid}
\end{table}

\subsection{Momentum dependence of particle identification performance}
\label{sec:momentum_dependence}

\begin{figure}[ht]
    \centering
    \includegraphics[width=0.60\linewidth]{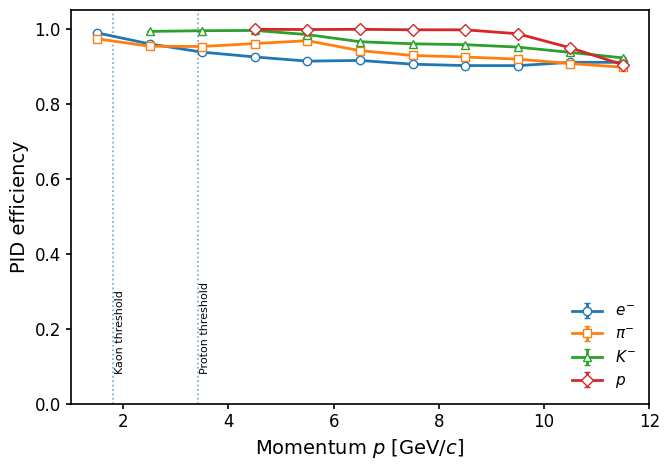}
    \caption{PID efficiency as a function of true particle momentum for the CNN+XGBoost classifier. The efficiencies are evaluated in common 1~GeV/$c$ momentum bins. Binomial statistical uncertainties are included and are generally smaller than the marker size. For each particle species, only bins whose full momentum interval lies above the corresponding Cherenkov emission threshold are shown. Bins below threshold or crossing the threshold are omitted from the plotted curves. The kaon and proton threshold positions are indicated by vertical dotted lines.}
    \label{fig:acc_vs_momentum}
\end{figure}

The momentum dependence of the classifier performance is essential for assessing the physics validity of the model. Figure~\ref{fig:acc_vs_momentum} shows the identification efficiency as a function of the true particle momentum for each species. The integrated efficiencies are 93.3\% for electrons, 94.1\% for pions, 96.6\% for kaons, and 98.3\% for protons.

The efficiency of identifying a given particle species $i$ is defined as
\begin{equation}
\varepsilon_i =
\frac{N(\hat{y}=i,\, y=i)}{N(y=i)}
= P(\hat{y}=i \mid y=i),
\qquad
i \in \{e,\ \pi,\ K,\ p\},
\end{equation}
where $y$ denotes the true particle species and $\hat{y}$ the species predicted by the classifier. The efficiency therefore quantifies the probability of correctly identifying a given particle $i$ in the presence of the competing hypotheses $\{e,\pi,K,p\}$.

\begin{figure}[ht]
    \centering
    \begin{minipage}{0.48\linewidth}
        \centering
        \includegraphics[width=\linewidth]{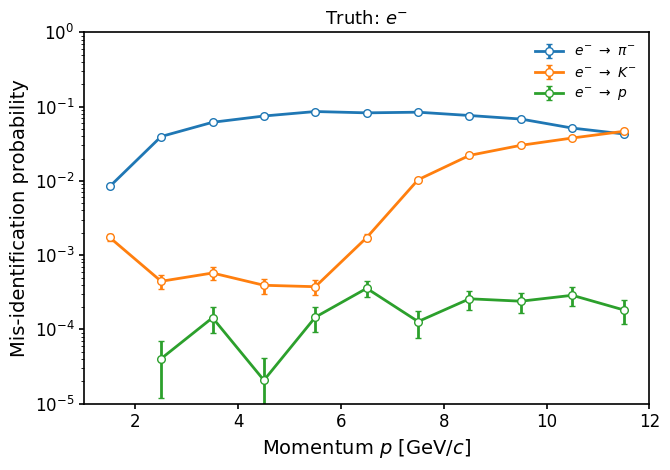}
        \vspace{0.1cm}
        (a)
    \end{minipage}
    \hfill
    \begin{minipage}{0.48\linewidth}
        \centering
        \includegraphics[width=\linewidth]{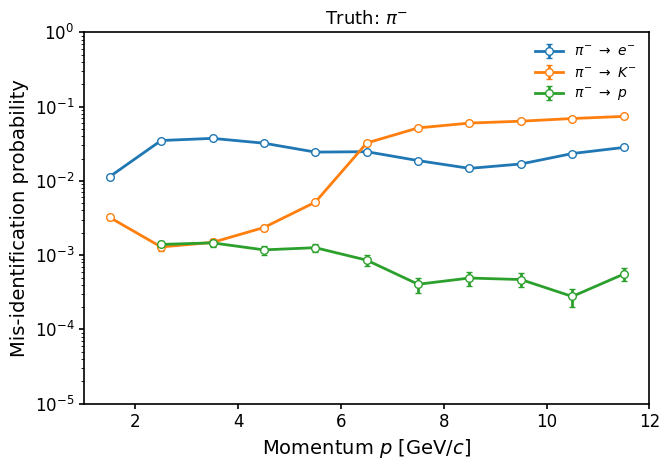}
        \vspace{0.1cm}
        (b)
    \end{minipage}

    \vspace{0.3cm}

    \begin{minipage}{0.48\linewidth}
        \centering
        \includegraphics[width=\linewidth]{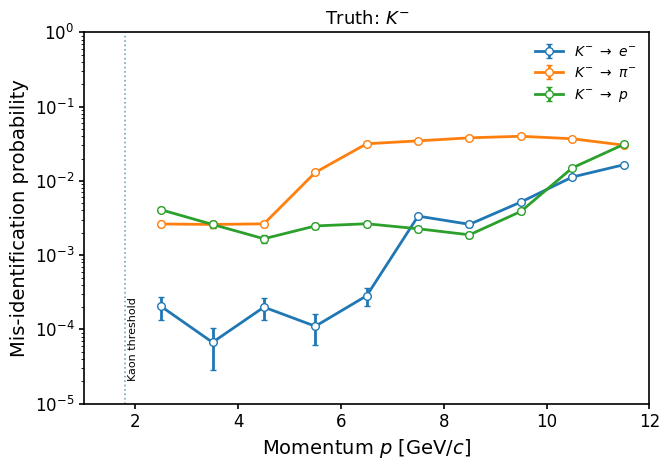}
        \vspace{0.1cm}
        (c)
    \end{minipage}
    \hfill
    \begin{minipage}{0.48\linewidth}
        \centering
        \includegraphics[width=\linewidth]{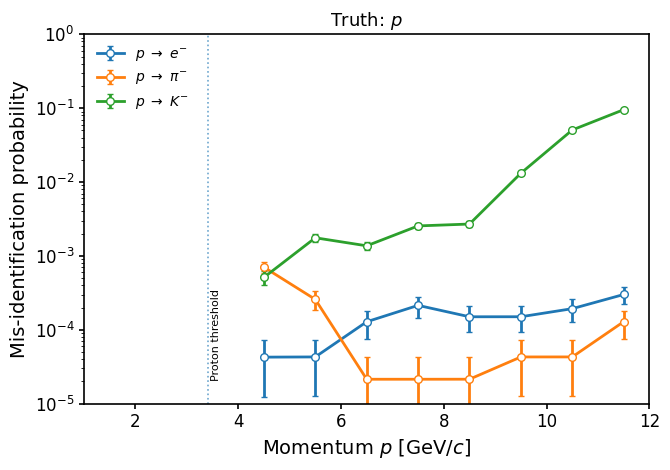}
        \vspace{0.1cm}
        (d)
    \end{minipage}

    \caption{Mis-identification probabilities as a function of true particle momentum for (a) electron, (b) pion, (c) kaon, and (d) proton truth samples. The probabilities are evaluated in common 1~GeV/$c$ momentum bins, with binomial statistical uncertainties shown as error bars. For each truth species, only bins whose full momentum interval lies above the corresponding Cherenkov emission threshold are shown. Bins below or crossing threshold are omitted from the plotted curves.}
    \label{fig:misid_vs_momentum}
\end{figure}

The mis-identification probabilities are presented in Fig.~\ref{fig:misid_vs_momentum} as a function of momentum for electron, pion, kaon, and proton truth samples. They are evaluated using the same common 1~GeV/$c$ momentum bins and threshold-aware selection as the PID efficiencies. For each truth species, bins whose full momentum interval lies below or crosses the corresponding Cherenkov threshold are omitted. Binomial statistical uncertainties are shown as error bars. For electrons, the dominant confusion channel is $e \to \pi$, which increases with momentum as Cherenkov angles converge in the ultra-relativistic regime, while $e \to K$ remains suppressed at low momentum and rises only at high momentum. Kaon mis-identification is primarily driven by confusion with pions at intermediate and high momentum, whereas the $K \to p$ channel remains subdominant. For pions, the leading confusion channel is $\pi \to K$, which increases with momentum, while $\pi \to e$ remains at the percent level and $\pi \to p$ is strongly suppressed. Overall, the mis-identification probabilities in this sanitized single-particle sample remain below the $10\%$ level up to 12~GeV/$c$. These high-momentum values are obtained under idealized conditions, however: as the Cherenkov angles converge in the ultra-relativistic regime the physical separation degrades as \rat{1}{p^2}, so the absence of chromatic dispersion, background, and occupancy in the present sample makes the quantitative high-momentum performance a best-case benchmark rather than a realistic in-situ estimate (see Section~\ref{sec:separation}).

\FloatBarrier
\subsection{Separation power}
\label{sec:separation}

The separation between two particle species $(i,j)$ is quantified using an effective score-separation metric, denoted $n_\sigma$. For a given momentum bin, this quantity is defined as
\begin{equation}
n_\sigma = \frac{|\mu_i - \mu_j|}{\sqrt{\sigma_i^2 + \sigma_j^2}},
\end{equation}
where $\mu_i$ and $\sigma_i$ are computed directly from the distribution of binary-classifier scores for species $i$ in that momentum bin. Dedicated binary XGBoost classifiers are trained separately for \rat{\pi}{K} and \rat{e}{\pi} separation to evaluate the momentum dependence of this effective separation metric. Because the classifier score is bounded between 0 and 1, the score distributions are not expected to be exactly Gaussian. We therefore interpret Eq.~(4.2) as an effective separation metric rather than as an exact Gaussian significance.

\begin{figure}[ht]
    \centering
    \includegraphics[width=0.65\linewidth]{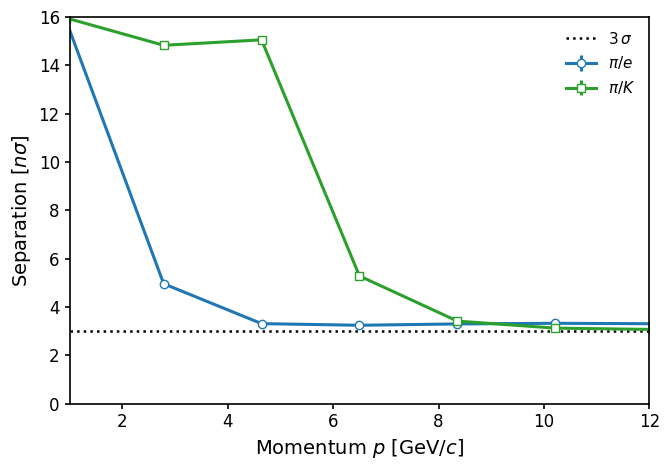}
    \vspace{0.3cm}
    \caption{Effective classifier-score separation $n_\sigma$ (Eq.~(4.2)) as a function of momentum for \rat{\pi}{K} and \rat{e}{\pi}, evaluated using dedicated binary classifiers. The quantity shown is a separation of the bounded binary-classifier scores, not the physical Cherenkov-angle significance, which degrades as \rat{1}{p^2} as the velocities converge ($\beta \to 1$). The plateau near the $3\sigma$ reference level (dotted line) at high momentum reflects the bounded-score metric evaluated under idealized single-particle conditions and should be read as a best-case benchmark; see Section~\ref{sec:separation}.}
    \label{fig:nsigma_combined}
\end{figure}

The \rat{e}{\pi} and \rat{\pi}{K} effective separation as a function of momentum is shown in Fig.~\ref{fig:nsigma_combined}. The separation is largest at low and intermediate momentum and decreases at high momentum as particle velocities converge. At the highest momenta studied, the effective separation approaches the $3\sigma$ reference level. We emphasize that $n_\sigma$ in Eq.~(4.2) is an effective separation of the bounded binary-classifier scores and is not equal to the physical Cherenkov-angle significance. The latter scales as $\Delta\theta_c \simeq \tfrac{\Delta(m^2)}{2\,n\,\theta_{\mathrm{sat}}\,p^2}$ once both species saturate, and therefore decreases as \rat{1}{p^2}; under realistic chromatic dispersion, emission-point and timing smearing, and background, the physical \rat{e}{\pi} separation falls below $3\sigma$ well within the momentum range studied. The plateau in Fig.~\ref{fig:nsigma_combined} reflects the idealized, sanitized single-particle conditions together with the bounded-score metric, and the quantitative values for $p \gtrsim 7$~GeV/$c$ should be read as a best-case benchmark. Within the pfRICH design range ($\lesssim 7$~GeV/$c$), the relevant hadron separation is comfortably above $3\sigma$, satisfying the design requirement.

\subsection{Feature importance analysis}
\label{sec:features}

\begin{figure}[h]
    \centering
    \includegraphics[width=0.65\linewidth]{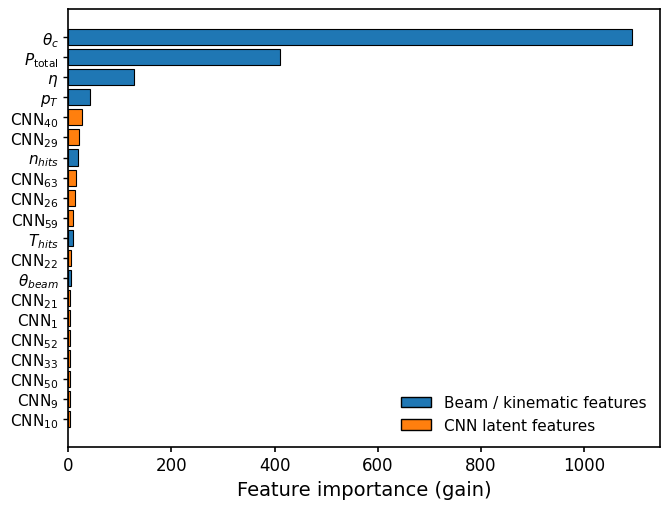}
    \caption{Feature-importance ranking for the hybrid CNN+XGBoost model. Features are color-coded by type: beam/kinematic features (blue) and CNN latent features (orange).}
    \label{fig:feature_importance}
\end{figure}

The feature-importance in Fig.~\ref{fig:feature_importance} shows the XGBoost classifier rankings for the hybrid model. The reconstructed Cherenkov angle $\theta_c$ dominates the ranking, consistent with its direct physical relationship to particle mass through Eq.~\eqref{eq:thetaapprox}. The total momentum $P_{\mathrm{total}}$ ranks second, followed by the pseudorapidity $\eta$ and transverse momentum $p_T$.

The feature-importance ranking shows that the dominant discrimination is provided by the reconstructed Cherenkov angle and momentum-related variables. Although some CNN latent variables appear in the ranked list, their contribution is subdominant compared with the leading tabular detector and kinematic features. As a cross-check, a baseline XGBoost classifier trained on the eight tabular features alone, without any CNN latent variables, give performance compatible with the full hybrid model on the same train/test split, with no significant difference between the two configurations. This is expected in a simplified dataset with little realistic noise, background, or overlapping-ring structure, where the reconstructed Cherenkov angle and kinematic variables already contain most of the PID information. We therefore treat the tabular XGBoost classifier as the operative baseline in this sample, and the CNN latent representation as an exploratory addition that does not improve performance here but is expected to contribute under realistic detector conditions. The CNN component is therefore retained as part of the hybrid proof-of-concept framework, with its potential impact to be further evaluated in future studies including realistic detector occupancy, backgrounds, and overlapping rings.

\section{Conclusion}
\label{sec:conclusion}

We have presented a hybrid machine-learning approach to enhance particle-identification (PID) performance for the proximity-focusing RICH detector of the ePIC experiment at the Electron–Ion Collider. Efficient hadron separation over a broad momentum range in the backward region is a key requirement for precision measurements of hadronic structure in electron–nucleus collisions.

Using a standalone \textsc{Geant4} simulation of the pfRICH, we have demonstrated that machine-learning techniques can significantly improve the reconstruction of Cherenkov patterns and the discrimination between pions, kaons, and protons across the relevant kinematic phase space. The proposed hybrid model, combining CNN-based pattern recognition with gradient-boosted decision trees, achieves \rat{\pi}{K} separation comfortably above $3\sigma$ across the pfRICH design range (up to 7~GeV/$c$) using a dedicated binary classifier, consistent with the pfRICH design requirement for hadron identification. The \rat{e}{\pi} separation is strong at low momentum but, as expected for an aerogel radiator, degrades once both species saturate; the high-momentum behavior reported here is obtained under idealized single-particle conditions and should be read as a best-case benchmark. The physical Cherenkov-angle separation degrades as \rat{1}{p^2} as the velocities converge, and a realistic estimate at high momentum requires the full detector simulation including chromatic dispersion, backgrounds, and occupancy, which we identify as the natural next step.

These results indicate that advanced ML-based PID methods can reinforce and extend the physics capabilities of the EIC~\cite{AbdulKhalek:2021gbh}. The approach is readily applicable to other ePIC sub-detector systems and provides a flexible framework for integrating image-based detector information with traditional kinematic observables.

As a next step, the model will be evaluated within the full ePIC software framework, including realistic background conditions. This will also enable a direct, like-for-like comparison of the present ML approach against a classical likelihood-based Cherenkov-angle reconstruction under realistic occupancy and background conditions. Future developments will also explore dedicated denoising techniques to further enhance signal quality and strengthen the PID performance of the pfRICH and related detector systems.

\section*{Acknowledgments}
We thank the ePIC collaboration for providing the pfRICH detector simulation framework and for valuable discussions related to particle-identification performance at the Electron--Ion Collider. We are grateful to the developers of the standalone Geant4 pfRICH simulation for making their tools publicly available.

This work is supported by the Simons Foundation. 

\bibliographystyle{JHEP}
\bibliography{biblio.bib}

\end{document}